\def\real{{\tt I\kern-.2em{R}}}%Typos corrected as of 8/8/00
\def\nat{{\tt I\kern-.2em{N}}}%not yet placed on archive as of 8/8/00

\def\realp#1{{\tt I\kern-.2em{R}}^#1}
\def\natp#1{{\tt I\kern-.2em{N}}^#1}
\def\hyper#1{\ ^*\kern-.2em{#1}}

\def\hyperrealp#1{{\tt ^*{I\kern-.2em{R}}}^#1} 

\def\hypernatp#1{{{^*{{\tt I\kern-.2em{N}}}}}^#1}

\def\leaderfill{\leaders\hbox to 1em{\hss.\hss}\hfill}
\def\srealp#1{{\rm I\kern-.2em{R}}^#1}

\def\pars{\par\smallskip}
\def\parm{\par\medskip}

\def\b#1{{\bf #1}}
\def\ref#1{$^{#1}$}

\def\m@th{\mathsurround=0pt}
\def\rightarrowfill{$\m@th \mathord- \mkern-6mu \cleaders\hbox{$\mkern-2mu 
\mathord- \mkern-2mu$}\hfil \mkern-6mu \mathord\rightarrow$}
\def\leftarrowfill{$\mathord\leftarrow
\mkern -6mu \m@th \mathord- \mkern-6mu \cleaders\hbox{$\mkern-2mu 
\mathord- \mkern-2mu$}\hfil $}
\def\noarrowfill{$\m@th \mathord- \mkern-6mu \cleaders\hbox{$\mkern-2mu 
\mathord- \mkern-2mu$}\hfil$}
\def\orgate{$\bigcirc \kern-.80em \lor$}
\def\andgate{$\bigcirc \kern-.80em \land$}
\def\inverter{$\bigcirc \kern-.80em \neg$}
\magnification=\magstep1 
\tolerance 10000
\baselineskip  14pt
\hoffset=0.25in
\hsize 6.00 true in
\vsize 8.75 true in

\def\id{\par\hangindent2\parindent\textindent}
\def\textindent#1{\indent\llap{#1}}  
\centerline{{\bf General Relativity and Contrary Predictions}\footnote*{This research is sponsored in part by a grant from the U. S. Naval Academy Research Council. Note: I am not at all certain about the results of this research. The results presented in this paper could be in error for various reasons. Although I have attempted to apply the procedures as outlined in various texts, if my calculations are correct, then my physical interpretations may need modification or may exhibit various errors.}}\medskip
\centerline{Robert A. Herrmann}\parm
\centerline{Mathematics Department}
\centerline{U. S. Naval Academy}
\centerline{572C Holloway Rd}
\centerline{Annapolis$,$ MD 21402-5002 USA}
\centerline{16 June 1998}\bigskip
\noindent {\it Abstract:} Using the accepted methods to extract natural system behavior from the Einstein-Hilbert gravitational field tensor equation$,$ a new coordinate transformation is analyzed. It is demonstrated  that these extraction methods yield specific contradictions. Two such contradictions are that for an electrically neutral (non-rotating) Schwarzschild configuration there is an infinite redshift surface that is not an event horizon and for a specific Schwarzschild configuration$,$ no dustlike particle external to this configuration will$,$ with respect to proper time$,$ ever gravitationally collapse radially to the surface of the configuration. \parm
\centerline{\bf Einstein's Equivalence Principle}\parm
  Within our universe$,$ is it possible to differentiate physically between the force effects of a gravitational field and the apparent force needed to produce a bodies accelerated motion$,$ where the only observations that can be made are behavior of  test bodies affected by these forces? Notice that this question is relative to properties associated with ``forces'' and not the equivalence of initial and gravitational mass. With the help of his famous ``elevator analogy,'' Einstein wrote:\par
{\leftskip=0.5in \rightskip=0.5in \noindent
[Let $K'$ be a system of reference such that] relative to $K'$ a mass sufficiently distant from other masses has an accelerated motion such that its acceleration and direction of acceleration are independent of its material composition and physical state. \par    
Does this permit an observer at rest relative to $K'$ to draw the conclusion that he is on a ``really'' accelerated system of reference? The answer is negative; for the above mentioned behavior of freely moving masses relative to $K'$ may be interpreted equally well in the following way. The system of reference $K'$ is unaccelerated$,$ but the space region being considered is under the sway of a gravitational field$,$ which generates the accelerated motion of the bodies relative to $K'.$ (Ohanian and Ruffini$,$ 1994$,$ p. 53)\par}\par   
However$,$ Fock wrote:\par
{\leftskip=0.5in \rightskip=0.5in \noindent As was mentioned$,$ Einstein considered that from the point of view of the Principle of Equivalence it is impossible to speak of absolute acceleration just as it is impossible to speak of absolute velocity. We consider this conclusion of Einstein's to be erroneous . . . (1959$,$ p. 208)\par}\par
\noindent Fock gives an example for his claim that uses a rotating  noninfinitesimal (i.e. non-local)  physical structure and that contradicts the above Einstein statement. It seems that under most conditions experienced within our universe  such effects$,$ for macroscopic entities$,$  can be differentiate one from another. This lead Fock and most theorists to modify$,$ at the least$,$ Einstein's original hypothesis so as not to forget ``that the nature of equivalence of fields of acceleration and of gravitation is strictly local.'' (Fock$,$ 1995$,$ p. 369)\par      
This is not the last word on this principle$,$ however. The mathematical model chosen to model the Einstein theory of gravity uses the infinitesimal calculus. The methods of corresponding physical behavior to the mathematical structure were specifically ignored. One needs to approximate physical infinitesimal measures before such a structure can be considered as a meaningful mathematical model. Gravitational {\it tidal effects} are the effects that a gravitational field has upon macroscopic physical objects. Because of the existence a special instrument called a {\it gravity gradiometer}$,$ an instrument that can measure the local differences in the tidal effects or what is termed the {\it tidal fields,} the principle requires a further adjustment. 
The gravity gradiometer can be reduced to a comparatively small size and$,$ since it can be so reduced$,$ it represents an approximate  physical ``infinitesimalizing process.'' Apparently$,$ as this instrument is reduced in size it is less likely to measure a differences between tidal forces and the forces associated with pure acceleration. This has lead Ohanian and Ruffini to make a further adjustment to this principle; an adjustment in which I concur. \par    
The following is still a general statement relative to all possible gravitational fields within our universe. For this theory$,$ the term {\it point particles} intuitively refers to physical entities that are very small$,$ but not small enough to be controlled by pure quantum mechanical properties. Further$,$ such entities are$,$ usually$,$ restricted to behavior within very small ``distances'' and are assumed not to have a significant gravitational field themselves so that they do not measurably influence the ``stronger'' gravitational field being investigated. It is the differential equation model that$,$ by a special type of summation of these effects$,$ allows one to describe behavior of point particles over macroscopic ``distances.'' \par
{\leftskip=0.5in \rightskip=0.5in \noindent Gravitation and acceleration are only equivalent as far as the translational motion of point particles is concerned (this amounts to what we call the Galileo principle of equivalence$,$ sometimes also called the ``{\it weak}'' principle of equivalence.) (Ohanian and Ruffini$,$ 1994$,$ p. 53)\par}\par    
The discussion in this section indicates how$,$ by means of further reflection or experimentation$,$ an original hypothesis is altered without substantially effecting a physical theory's conclusions. However$,$ with respect to the Einstein's General Theory$,$ the following sections of this article  demonstrate that some of the most basic hypotheses used to deduce physical conclusions may need to be altered in order to avoid absolute contradictions. \parm   
\centerline{{\bf Further Basic Hypotheses}}\parm
  Einstein rejected the concept of the {\it privileged observer.} Intuitively$,$ this means an actual entity that can take measurements and represent these measurements in a specific manner. Further$,$ a privileged observer would describe a privileged form of physical law by representing physical law only in this specific manner.
Specifically$,$ Einstein rejected:\par
{\leftskip=0.5in \rightskip=0.5in \noindent . . . the Newtonian concept of a privileged observer$,$ at rest in absolute space. . . (Lawden$,$ 1982$,$ p. 127)\par}\par
\noindent This concept also is intended to imply  that there is no privileged position within our universe from which to observe. In order to 
model mathematically the ``no privileged observer'' concept$,$ we need to model for this Einstein theory the concept of ``observer.'' \par
{\leftskip=0.5in \rightskip=0.5in \noindent
(1) An observer is modeled by a specific coordinate system$,$ where a coordinate system is a definable process that will measure each member of an ``ordered 4-tuple'' $(x_1,x_2,x_3,x_4).$ \par}\par   
Thus whenever you read the term ``observer'' it signifies mathematically 
``a specific coordinate system.'' But not all definable process are allowed in the definition of what constitutes a coordinate system. And$,$ on the other hand$,$ the concept of ``definable'' depends intuitively  upon the language and logic one applies. So we cannot actually known what is or is not ``definable.''  But$,$ physically$,$ it is usually required that for any two intuitively different physical events$,$ at least one of the members of the two representing 4-tuples is different. The difference in 4-tuple coordinates depends upon the objects used. Usually$,$ the coordinates themselves are real or complex numbers and we confine these remarks to this case. \par    
In the physical sciences$,$ coordinate measures for physical events  are used in an attempt to glean from the {\it relations between} these measures what has been termed as natural law. Under the assumption of no privileged observer$,$ if two different modes of coordinate measure as modeled by the concept of ``coordinate systems'' are used to express relations between the 4-tuples that represent different events$,$ then one should be able to  glean from these relations the same physical law. As mentioned$,$ in order to model this idea$,$ the concept of the no physically preferred coordinate system is introduced. Ohanian (1976$,$ p. 253) states this as follows:\par
{\leftskip=0.5in \rightskip=0.5in \noindent It must be possible to express physical laws in {\it any} coordinate system. This is obvious since ultimately these laws express nothing but {\it relationships} between spacetime coincidences. The result of any experiment can always be reduced to a statement of the form `. . . when the ends of the two wires coming out of instrument A are made to {\it coincide} with the two terminals of instrument B$,$ then the pointer on the face of instrument A will coincide with the third mark on the scale$,$ etc.,'  that makes no reference to 
coordinates.\par}\par   
There are actually infinitely many mathematical theories where equations can be considered  invariant in form under coordinate transformations. For example$,$ consider the expression $A^{uv},$ where $u$ and $v$ vary independently from $1,\ldots, 4.$ Suppose that these 16 ``components'' are expressed in terms of real valued 
variables $x_i,$ $1 \leq i\leq 4,$ and further suppose that we have four other variables $\overline{x_i},\ 1\leq i\leq 4$ and that these new variables are related to the original ones by four real valued equations $\overline{x_i} = f_i(x_1,x_2,x_3,x_4),\ 1\leq i\leq 4.$  Consider a fixed relation $R=R(\overline{x_i},x_i)$ defined in terms of the 8 variables. Require that the only acceptable objects that can be used must have the following transformation law. That when we substitute for all the $x_i$ variables that appear in  
$R$ and $A^{uv}$$,$ the transformed result $\overline{A}^{uv}$ satisfies the rule $\overline{A}^{uv} = RA^{uv}$. It follows easily that if we have three such collections of components $A^{uv},\ B^{uv},\ C^{uv}$$,$ then  equations such as $A^{uv} = B^{uv} +C^{uv}$ and $C^{uv}B^{uv} = B^{uv}C^{uv}$ when they are ``transformed'' to the variables $\overline{x_i}$ have the form
$\overline{A}^{uv} = \overline{B}^{uv} +\overline{C}^{uv}$ and $ \overline{C}^{uv}\overline{B}^{uv}= \overline{B}^{uv}\overline{C}^{uv}.$ \par    
Since the concepts of the calculus and differential equations are used in the Einstein theory$,$ the transformation law $R$ that is used for the components is based upon how one usually writes the differential of each of the new variables. Due to this and other concerns$,$ 
the mathematical model used does not satisfy the idea of 
{\it any} coordinate system$,$ but restricts such systems to the set of all {\it proper} coordinate systems. \par    
In the Einstein theory$,$ the proper coordinate systems are related to two different modes of coordinate measurement used  to characterize the same physical event by what appear to be two different 4-tuples. The two different systems used should be related so that the two different 4-tuples can be associated uniquely one with another and this association would indicate that the physical event is the same event but is simply being observed in different ways. In order to model mathematically this concept$,$ the idea of the {\it proper coordinate transformation} is used.\par
{\leftskip=0.5in \rightskip=0.5in \noindent 
(2) For the 4-tuple's$,$ proper coordinate systems are related by four equations$,$ called {\it coordinate transformations} so that$,$ at least within a local real spacetime region$,$ the coordinates of one of the 4-tuples can be used to calculate uniquely the coordinates of the other 4-tuple$,$ and conversely.\par}\par   
But can one actually write down all of the coordinate transformations that satisfy (2)? The answer is no since there are infinitely many of them. In the Einstein theory$,$ the actual coordinate systems used to glean physical behavior are obtained by ``trial and error'' (Misner$,$ Thorne and Wheeler$,$ 1973$,$ p. 832). Further$,$ in order to have a structure that might be used to ``express physical laws in {\it any} coordinate system,'' although the actual set is a restriction to proper systems as mentioned$,$  Einstein selected the {\it absolute  differential calculus} or {\it tensor analysis} (Ricci-Curbastro and Levi-Civita$,$ 1901) as the appropriate mathematical structure.  The use of the absolute differential calculus$,$ however$,$ further restricts the proper coordinate systems to those that can be transformed by means of {\it regular coordinate transformations}. \par
{\leftskip=0.5in \rightskip=0.5in \noindent 
(3) We add to (2) the requirement that the four transformation equations must satisfy a special set of partial differential expressions$,$ at the least$,$ in a region called a neighborhood ``about'' a specific 4-tuple.\par}\par   
In order to minimize the technical aspects of this article$,$ the relations required by (3) will not be specifically stated. Fock (1956) argued that there would need to be an additional constraint placed upon the coordinates used. His work incorporates what are called the {\it harmonic coordinate conditions} and he shows that under this constraint there is$,$ apart from a Lorentz transformation$,$ a privileged (unique) coordinate system called the {\it harmonic} system. However$,$ Misner et. al. (1973)
reject this contention.\par
{\leftskip 0.5in \rightskip 0.5in \noindent . . . to make a choice among coordinate systems is exactly what the geometrodynamic law cannot and must not have the power to do. (p. 409)\par}\par
 \noindent Indeed$,$ Ohanian and Ruffini (1994) also make the same observation.  Note that$,$ for reference purposes$,$ some quotations that follow will be denoted by numerical markers $\{ \cdot \}.$\par
{\leftskip 0.5in \rightskip 0.5in \noindent Of course$,$ it is possible to change the {\it form} of any [line element] solution by a coordinate transformation. $\{\b 1\}$ Therefore$,$ two solutions will be said to agree if they differ by no more than a coordinate transformation; such solutions are physically identical. $\{\b 1\}$ (p. 397)\footnote*{Statement $\{\b 1\}$ is demonstrated as false by Ohanian and Ruffini when they introduce a physical interpretation for the Kruskal-Szekeres coordinate system. Other technical requirements or different interpretation techniques may be necessary. Different coordinate systems can lead to different ``physical'' interpretations and even ``new physics.'' If found to be valid, the results in this article also demonstrate that contrary ``physics'' can result from different coordinate transformations.} \par}\par    
In ordinary non-coordinate geometry$,$ there are certain intuitive concepts that appear to be related. No coordinate system$,$ as such$,$ should alter these geometric relations. One application of the absolute differential calculus allows some concepts from differential geometry to be stated in a form that is independent from coordinate systems of type (3). For example$,$ under the restrictions stated above$,$ if all components of the Riemann-curvature tensor  are zero (i.e. $R^\alpha_{\;\; \beta\mu\nu}= 0$) throughout all of geometric space$,$ then the space is Euclidean (i.e. flat) and distances can be measured by a simple Pythagorean relation. This flatness is thought to be independent of the coordinate system used and$,$ hence$,$ we have this tensor expression. This mathematical structure was devised so that the {\it form} of this equation does not change when coordinate transformation of the type (3)$,$ with the special transformation law $R,$ are applied. Apparently$,$ this is the basic reason Einstein selected this structure as the one that would model physical law and preserve the no privileged observer hypothesis. Further$,$ he could use$,$ as a model$,$     
the previous language of Riemannian geometry with its reliance upon differential equation solutions to geometric problems. However$,$ confusion lies in the fact that Einstein's theory is but a physical {\it model} that predicts the behavior of physical entities but does not specifically define any physical entity that comprises space[time]. Patton and Wheeler (1975) state:\par
{\leftskip 0.5in \rightskip 0.5in \noindent Five bits of evidence argue that geometry is as far from giving an understanding of space as elasticity is from giving a understanding of a solid. . . . Tied to the paradox of the Big Bang and collapse is the question$,$ what is the substance out of which the universe is made? (p. 539$,$ 543)\par}\parm       
\centerline{\bf The Major Hypothesis?}\parm
  For purposes of simplicity$,$ in all that follows$,$ ``light units'' will be used for our measurements. That is: we set the famous constant $c =1.$ [Note: due to the use of light units$,$ if a unit analysis is applied to the following expressions in this paper$,$ then the unit statements will appear to be inconsistent.] Einstein's General theory is a generalization of his Special theory. For his Special theory$,$ it was observed by Minkowski that the famous Lorentz transformation equations expressed in terms of $t,\ x,\ y,\ z$ and $t',\ x',\ y',\ z'$ coordinates$,$ which are regular$,$ satisfy the differential expression
$$dt^2 - (dx^2 +dy^2 + dz^2) = (dt')^2 -((dx')^2 + (dy')^2) + (dz')^2).\eqno (1)$$
(Although it will not be discussion further$,$ I mention that this equation can also be arrived at by considering the behavior of electromagnetic radiation. Indeed$,$ there is a derivation from first-principles that shows explicitly that the variables in (1) are actually physical measures made by ``infinitesimal'' light-clocks and should not be considered as any other possible mode  of measurement if consistency is to be maintained (Herrmann,1994b).)\par    
The expression (1) is denoted by $ds^2$ and is called the spacetime {\it interval} or the more descriptive infinitesimal {\it Chronotopic interval}. Further$,$ (1) is called a ``metric'' but$,$ so as not to confuse this term with that used in other mathematical theories$,$ it will be termed$,$ in all the follows$,$ as the Minkowski {\it line element.} Now the equality in (1) holds for those many  transformations that comprise the ``Lorentz ( or Poincar\'e) group.''\par    
In four-dimensional differential geometry$,$ we have a similar expression for {any} four variables $x_1,x_2,x_3,x_4.$ 
$$g_{11}dx_1^2 + g_{12}dx_1\,dx_2+ \cdots + g_{34}dx_3\, dx_4 + g_{44}dx_4^2, \eqno (2)$$
where the line element coefficient (i.e. metric coefficients $g_{\mu\nu}$) subscripts vary independently over the four number set $\{1,2,3,4\}$$,$ the differentials of the variables also vary in the same manner as indicated$,$ and the expressions are multiplied and added as indicated. As an example$,$ the Minkowski line element has $g_{11} = 1,\ g_{22}=g_{33}=g_{44}=-1$ and all others are zero. This produces$,$ in this case$,$ the spacetime {\it signature} $(+,-,-,-)$. Expression (2)$,$ with its 16 terms$,$ is also denoted by $ds^2$ and called the {\it general line element.}
Further$,$ Einstein summation notation is used. In this form,
$$ds^2 = g_{\mu\nu}dx_\mu\, dx_\nu,\eqno (3)$$ represents$,$ first$,$ the summands$,$ where the $\mu$ and $\nu$ are thought of as varying over all the ordered pairs formed from the set $\{1,2,3,4\}$ and$,$ secondly$,$ they are all added together. \par   
The assumed major hypothesis is not dependent upon whether one associates the $g_{\mu\nu}$ with terms taken from geometry or not. This assumed major physical hypothesis has been alluded to by Ohanian and Ruffini (1994) in the $\{\b 1\}$ sentence from the last section. To explain in detail this hypothesis$,$ I quote from two highly distinguished sources$,$ sources that differ greatly$,$ however$,$ philosophically.\par
{\leftskip=0.5in \rightskip=0.5in \noindent
This [Equation (3)] represents all possible mechanical systems described in terms of all possible co-ordinate systems [proper]. If the coefficients are all given specific values,
then the equation represents a particular mechanical system described in terms of a particular system of co-ordinates. $\{\b 2\}$ Any change in this expression for $ds^2$ which may be brought about by a mathematically permissible transformation of co-ordinates represents the same mechanical system differently described: $\{\b 2\}$ (Dingle$,$ 1950$,$ p. 85) \par}\pars
{\leftskip=0.5in \rightskip=0.5in \noindent If a prediction is to be made of the geometry$,$ how much information has to be supplied for this purpose? The geometry of spacetime is described by the metric 
$$ds^2 = g_{\alpha\beta}({\cal P}) dx^\alpha\,dx^\beta;$$
of location $\cal P$
in spacetime. It might then seem that ten functions must be predicted; and$,$ if so$,$ that one needs for the task ten equations. Not so. $\{\b 3\}$ Introduce a new set of coordinates $x^{\overline{\mu}}$ by way of the coordinate transformations
$$x^\alpha = x^\alpha(x^{\overline{\mu}}),$$
and find the same spacetime geometry$,$ with the same bumps$,$ rills$,$ and waves$,$ described by an entirely new set of metric coefficients $g_{\overline{\alpha}\overline{\beta}}.$ $\{\b 3\}$ (Misner$,$ et. el. 1973 p. 408) \par}\pars  
  We then have a general reason why the tensor structure is used for the classical theory.\par
{\leftskip=0.5in \rightskip=0.5in \noindent The gravitational theory of Einstein is based upon the following postulate:
{\it Principle of general invariance: All laws of physics must be invariant under general coordinate transformations.} (Ohanian and Ruffini$,$ 1994$,$ p. 374)\footnote*{As pointed out in the last footnote$,$ this statement$,$ $\{\b 1\},$ $\{\b 2\},$ and $\{\b 3\}$ (the major hypothesis) must be interpreted very carefully and may be false.} \par}\parm  
\centerline{\bf The Coordinate Method}\parm
  The Einstein-Hilbert gravitational field equation$,$ without the cosmological constant$,$ is expressed in tensor form as 
$$R_{\mu\nu} -{{1}\over{2}}g_{\mu\nu}R =-8\pi GT_{\mu\nu},\eqno (4)$$
where the tensor $R_{\mu\nu}$ and scalar $R$ on the left side are expressed completely in terms of the $g_{\mu\nu}$ and their derivatives$,$ and the components  on the right expressed by the tensor $T_{\mu\nu}$ are components of the physical ``energy-momentum'' tensor. The $G$ is the universal constant of gravitation. It is now obvious why a geometric language is used as a model since all the corresponding terminology from differential geometry that involves the $g_{\mu\nu}$ may be utilized. \par    
Our major concern is a solution to (4) in the exterior regions of a universe that contains but one centrally symmetric homogeneous nonrotating  distribution of matter. We begin this investigation with the following general result$,$ where we have substituted $\rho$ for the variable $r$.\par
{\leftskip=0.5in \rightskip=0.5in \noindent If we use ``spherical'' space coordinates $\rho,\ \theta,\ \phi,$ then the must general centrally symmetric expression for $ds^2$ is 
$$ds^2 = h(\rho,t)dr^2 +k(\rho,t)(\sin^2\theta\ d\phi^2 +d\theta^2) +l(\rho,t) dt^2 +a(\rho,t)d\rho\, dt,\eqno (*)$$
where $a,\ h,\ k,\ l$ are certain functions of the ``radius vector'' $\rho$ and ``time'' $t.$ But because of the arbitrariness in the choice of a reference system in the general theory of relativity$,$ we can still subject the coordinates to any transformation that does not destroy the central symmetry of $ds^2$; this means that we can transform the coordinates $\rho$ and $t$ according to the formulas
$$\rho = f_1(r',t'),\ t = f_2(r',t'),$$ where $f_1,\ f_2$ are any functions of new coordinates $r',\ t'.$'' (Landau and Lifshitz$,$ 1962$,$ p. 324)\par}\par   
I point out that such transformations are often confused since$,$ after the substitutions are made$,$ the ``prime'' notation is often dropped from the new expression. The standard procedure now continues in the following manner.
One usually choices first$,$ $a(\rho,t)=0,$ Then $k(\rho,t)$ can be chosen so that it appears as $-k'(\rho,t)\rho^2,$ where $k'(\rho,t)$ is to be determined. When $k'(\rho,t) =1,$ the usual $\theta, \ \phi$ portion of the standard spherical coordinate transformation is obtained. The gravitational field is one produced by a centrally symmetric homogeneous nonrotating distribution of matter.\par    
Since physically this line element is a generalization of equation (1)$,$ a choice for $k'(\rho,t)$ is made so that the line element will become 
{\it asymptotically flat}. This means that as $\rho \to \infty$ this new element will correspond to   
the Minkowski line element as it is 
expressed in spherical coordinates. Application of this {\it boundary condition} implies that  
$k'(\rho,t) \to 1,$ as $\rho \to\infty.$
Similar restrictions are applied to functions $h$ and $l.$ Hence$,$ letting $x_1 = t,\ x_2 = \rho,\ x_3 = \theta,\  x_4 = \phi$$,$ we have that 
$g_{11} = l(\rho,t),\ g_{22}=h(\rho,t),\ g_{33}= -k'(\rho,t)\rho^2,\ g_{44}= 
-k'(\rho,t)\rho^2\sin^2\theta$ and all other $g_{\mu\nu} = 0.$\par   
  For the physical electrically neutral case considered here (the Schwarzschild configuration)$,$ where we include the static condition that the field is time independent and time symmetric (i.e. unchanged by time reversal)$,$ Schwarzschild (1916)  found the following values for the functions and$,$ hence$,$ for the corresponding $g_{\mu\nu}$ that satisfy the Einstein-Hilbert equation for the gravitational field external to the mass and that satisfy the ``Minkowski condition at infinity.'' Let $k'(\rho,t) = 1,\ g_{11} = (1-r_s/\rho),\ g(22) = -(1-r_s/\rho)^{-1},$ where $r_s = 2GM,$ $G$ is the universal gravitational constant$,$ and $M$ the mass of the object. Obviously$,$ this Schwarzschild line element is not defined at certain physically meaningful locations. In particular$,$ when $\rho = r_s$ and $\rho =0$ and$,$ hence$,$ the geometry ``. . . appears to behave badly. . .''(Misner$,$ et. al$,$ 1973$,$ p. 820) near these points. However$,$ the standard claim is that these values for the $g_{\mu\nu}$ satisfy the Einstein-Hilbert equation for $\rho> r_s$ and for positive $\rho < r_s.$ (Lawden$,$ 1982$,$ p. 155) Further$,$ if the actual Schwarzschild configuration has $\rho > r_s,$ then the line element only applies for a vacuum condition exterior to the configuration and any analysis for $\rho < r_s$ as no physical meaning. (Lawden$,$ 1982$,$ p. 146) Also$,$ Lema\^itre (1933) claims that the ``Schwarzschild singularity'' at $\rho = r_s$ is not a physical singularity since$,$ following the above major hypothesis on coordinate transformations$,$ it can  be removed by a coordinate transformation. 
Thus it is claimed that it is but a {\it coordinate singularity.}\par    
 Ohanian and Ruffini are emphatic when they write$,$ 
``It is important to recognize that the Schwarzschild `singularity'
at $\rho = r_s $ is not a physical singularity. The `singularity' in 
 Eqs. (2) and (3) is spurious$,$ or a pseudosingularity; it arises from an inappropriate choice of coordinates and can be eliminated by a change of coordinates.'' (Ohanian and Ruffini$,$ 1994 p. 440) Lema\^itre's transformation and all other coordinate transformations that have appeared since 1933 in accepted treatises on this subject claim that for this explicit physical scenario $\rho =0$ is a {\it physical singularity}
and is not simply a coordinate singularity. A physical singularity is defined
descriptively in various ways such as {\it spacetime behaves in an unusual manner} or {\it the theory brakes down completely} or {\it the known laws of nature are suspended} and other such statements. It also has a lengthy technical definition. Although $\rho =r_s$ represents what is claimed to be but a  coordinate singularity$,$ this does not preclude further analysis since spacetime may still have some interesting and unusual properties alone the Schwarzschild surface $\rho = r_s.$ The first basic analysis assumes that the Schwarzschild configuration is interior to $\rho = r_s$ where$,$ except on this surface$,$ the metric coefficients satisfy the Einstein-Hilbert equation exterior to the configuration. What method has been used to glean physical statements from line elements that only involve statements associated with coordinates and statements that appear to be relative only to the local (infinitesimal) environment?\parm
\centerline{\bf The Classical Global Method}\parm
  Ohanian and Ruffini (1994) state specifically the physical interpretation for the geometric concept called the {\it geodesic.}\par
{\leftskip=0.5in \rightskip=0.5in \noindent $\{\b 4\}$ Furthermore$,$ the motion of a particle in the field of the gravitational mass can be interpreted  as a free motion along the geodesic (p. 164) . . . . the geodesics of curved spacetime are worldlines of freely falling particles. $\{\b 4\}$ (p. 332)\par}\par   
Further,
Misner$,$ Thorne and Wheeler (1973$,$ p. 334) state that $\{\b 5\}$ ``Curvature is the simplest local measure of geometric properties.'' $\{\b 5\}$ Specifically$,$ the Riemann-curvature tensor yields the behavior of the tidal forces. The classical global method is to investigate the behavior of the Riemann-curvature tensor as a test particle in free-fall approaches various positions within the Schwarzschild field. Such a free-fall may be characterized at a point by a special set of coordinates called the {\it geodesic coordinates}. Using this technique$,$ it is found after some tedious calculations 
(Misner et. al. 1973$,$ p. 820-823; Ohanian and Ruffini$,$ 1994$,$ p. 440-441)$,$ using the Schwarzschild coordinates$,$ that 
each of the components of the Riemann tensor at points on the 
Schwarzschild surface $\rho = r_s$ is finite. This is stated physically by Ohanian and Ruffini (1994$,$ p. 440) as meaning that ``An astronaut falling through and crossing the surface $\rho= r_s$ will not feel anything unusual'' and this is considered as a ``local event.'' On the other hand$,$ the same type of analysis shows at least one of the Riemann-curvature components is unbounded at $\rho = 0$ and this local event is described as meaning that as an astronaut approaches this position the tidal force will destroy the individual. \par    
For reference purposes$,$ a singularity that results in unbounded tidal forces is called a {\it D-type} singularity. There are significant reasons way a D-type physical singularity needs a type of external shield. Without such a shield a D-type singularity$,$ if it actually forms due to gravitational collapse$,$ is called a ``naked singularity.'' \par
{\leftskip=0.5in \rightskip=0.5in \noindent . . . the formation of a naked singularity during collapse would be a disaster for general relativity theory. In this situation$,$ one cannot say anything precise about the future evolution of any region of space containing the singularity since new information could emerge from it in a completely arbitrary way. (Shapiro and
Teukolsky$,$ 1991$,$ p. 994)\par}\par  
 The usual name given for the shield that would surround a D-type singularity
is called the {\it event horizon}. \par
{\leftskip=0.5in \rightskip=0.5in \noindent Although the region $\rho <r_s$ [$\rho =r$] has no unusual properties of a local kind (except at $\rho =0$$,$ where there is a singularity)$,$ it does have some unusual properties of a global kind. As we will see from a careful analysis of the spacetime geometry$,$ the region $\rho <r_s$ is a {\it black hole.} By this is meant that no signal of any kind can emerge from the region $\rho <r_s$ and reach the region $\rho >r_s.$ The surface $\rho = r_s$ (regarded as a surface in spacetime . . . ) is the boundary between spacetime points that will become observable at some time$,$ and those that will never be observable by outsiders. This boundary of the black hole is called the {\it event horizon}. . . . The surface $\rho =r_s$ acts as a ``one-way membrane,'' through which signals can be sent in$,$ but not out. This is a global (or nonlocal) property because in order to test it$,$ we must examine the propagation of light signals and other signals and check what happens to them in the long run. (Ohanian and Ruffini$,$ 1994$,$ p. 443-444)\par}\par    
The event horizon is often confused with a surface of {\it infinite 
redshift} and they need not be the same (Ohanian and Ruffini$,$ 1994$,$ p. 438$,$ 466-468). But$,$ they coincide for the Schwarzschild field  as expressed by the following line element (5) discussed previously.
$$ds^2 = (1-r_s/\rho)dt^2 -(1-r_s/\rho)^{-1}d\rho^2 - \rho^2(d\theta^2 +\sin^2\theta\, d\phi^2).\eqno (5)$$
One way to determine$,$ with respect to coordinate measures $(\rho,t),$ the existence of infinite redshift surfaces is to consider $g_{11} =0$. 
As Ohanian and Ruffini (1994$,$ p. 464) state ``The vanishing of $g_{11}$ merely tells us that a particle cannot be at rest (with $d\rho=d\theta=d\phi = 0$) at these surfaces; only a light signal emitted in the radial direction can be at rest.'' This surface can also be determine by the discussion of 
``clocks.'' In all that follows$,$ $r_s$ has been substituted for $2GM$. \par 
{\leftskip=0.5in \rightskip=0.5in \noindent The surface $\rho=r_s$ is a surface of {\it infinite redshift.} A clock placed at rest near to $\rho = r_s$ shows a proper time
$d\tau = \sqrt{1-r_s/\rho}\, dt$ which approaches zero as $\rho \to r_s$; that is the clock runs infinitely slow compared with a clock at a larger distance. (Ohanian and Ruffini$,$ 1994$,$ p. 439)\par}\par   
The simplest way to analyze this event horizon is to consider but  
radial ``light'' signals. This is done in Ohanian and Ruffini (1994$,$ p. 445) by setting $ds^2= 0,\ d\theta = 0,\ d\phi = 0.$ Under normal conditions$,$ one would take $d\rho/dt < 0$ for light signals moving towards the Schwarzschild surface from a far distance and towards the center$,$ and  $d\rho/dt > 0$ for light signals moving from the center outward towards the surface. Note that Ohanian and Ruffini$,$ (1994$,$ p. 466) state that to find possible event horizons one puts
$d\rho/dt=0.$ 
Equation (5) yields two results$,$ in terms of the $(\rho,t)$ coordinate measures$,$ $d\rho/dt =\pm (1-r_s/\rho).$ A very useful procedure is the  ``light-cone'' analysis (Ohanian and Ruffini$,$ 1994$,$ p. 446).
 External to the Schwarzschild surface$,$ the signature of spacetime remains $(+,-,-,-)$. If the photon could pass the Schwarzschild surface$,$ then a change in signature takes place in the interior region. The value $d\rho/dt$ now becomes greater than zero. The interior signature of spacetime changes to $(-,+,-,-)$ for radially moving photons and the character of the light-cone diagrams change. The light-cones are orientated in a complementary direction throughout
the interior region$,$ and this orientation is interpreted to mean that$,$ at the least$,$ electromagnetic signals cannot leave the region interior to the Schwarzschild surface.  The same type of conclusion is reached if one investigates the behavior of ``material'' particles.
Also note that if the interior Schwarzschild field were extended to infinity$,$ then the corresponding Minkowski line element would have $g_{11}= -1$ and $g_{22} = +1$. It would seem as if  ``time'' and ``distance'' reverse their character with respect to radial movement. Indeed$,$ coordinate arguments relative to ``light'' signals that lead to the conclusion that a surface is an event horizon use this alteration in the character of spacetime and corresponding light-cone analysis.  \par    
 In order to investigate the possibility within General Relativity (GR) that photons and particles can actually pass through the Schwarzschild surface in one direction$,$ what are called ``well-behaved'' coordinate systems are introduced$,$ coordinate systems that lead to line elements that can be defined at $\rho=r_s$ and retain the D-type singularities for the transformed $\rho = 0.$ One of the less unusual transformations is the Eddington-Finkelstein transformation (EFT)
(Misner et. al. 1973$,$ p. 828-830). However$,$ the EFT is not defined at $\rho=r_s$ and$,$ indeed$,$ diverges there. Another is the  Kruskal-Szekeres transformation (KST) (Misner$,$ et. al. 1973$,$ p. 831-835; Ohanian and Ruffini 1994$,$ p. 449-459). Once again the KST is not defined at $\rho=r_s$ since it is not regular at that point. \par    
 In the EFT$,$ a new $\rho*$ is defined in terms of the original $\rho$ and $r_s,$ and a new ``time-like'' coordinate $V=t +\rho*$. [Notice that there are no compatible units for this $V.$] In the KST$,$ two coordinates are defined. One coordinate $v$ is defined in terms of the original $\rho,\ t$ and $r_s,$      
and a second coordinate $u$ in teams of the same entities. Although Rindler (1977$,$ p. 160) declares that $\{\b 6\}$ ``We must regard $x$ as a {\it radial} coordinate,'' $\{\b 6\}$ where Randler's $x$ is the $u,$ both of these new coordinates are dimensionless.
Further$,$ the same coordinate free description can be given relative to the ``relationship'' between the behavior of photons 
``within'' the transformed Schwarzschild surface and the D-type singularities. But$,$ since the line element may now be defined at what is the transformed Schwarzschild surface$,$ this additional aspect$,$ that could not be analyzed in Schwarzschild coordinates$,$ may be investigated. The investigation leads to the additional statement that ``photons can actually pass through the Schwarzschild surface in one direction and towards the center of the mass.'' The KST coordinates also add other major features that appear to be missing when viewed in the Schwarzschild system. 
This is explained as follows:\par
{\leftskip=0.5in \rightskip=0.5in \noindent How can this be?  . . . The answer must be that the Schwarzschild coordinates cover only a part of the spacetime manifold; they must be only a local coordinate patch on the full manifold. Somehow$,$ by means of the coordinate transformation that leads to the Kruskal-Szekers coordinates$,$ one has analytically extended the limited Schwarzschild solution for the metric to cover all (or more nearly all) of the manifold (Misner et. al. 1973$,$ p. 833)\par}\par
   
The above contention is illustrated by means of the well-known KST coordinate diagram where various $t$ and $r$ coordinates are graphed within the KST system (Misner et. al. 1973$,$ p. 834; Ohanian and Ruffini$,$ 1994$,$ p. 452). The  diagram is partitioned into sections I$,$ II$,$ III$,$ IV. This KST spacetime is said to be {\it maximal} and this is interpreted to mean that physical particles can appear and disappear only at physical singularities. The Schwarzschild solution only occupies or covers the regions I$,$ II.  
The physical descriptions for the KST coordinates is what has inspired many science fiction stories. The full KST space is the object that has introduced the ``white hole'' and ``Einstein-Rosen bridge (i.e. wormhole)'' into a cosmology. However$,$ one scientist writes: \par
{\leftskip=0.5in \rightskip=0.5in \noindent Through Kruskal's work is undoubtedly of high theoretic interest$,$ does it have practical application? At present$,$ perhaps not. Kruskal space would have to be {\it created in toto}: . . . . There is no evidence that full Kruskal spaces exist in nature. (Rindler$,$ 1977$,$ p. 164)\par}\par   
 As mentioned$,$ Fock (1959$,$ p. 194)  shows  that there does exist a unique coordinate system if you require the harmonic restriction.  Misner et. al (1973) reject this additional condition. However$,$ one can use a regular coordinate transformation and obtain Fock's line element.
Simply let $\alpha = r_s/2$ and then $\rho = r + \alpha.$ When this is substituted into the Schwarzschild equation (5)$,$ one obtains\medskip
$$ds^2 = \left({{r-\alpha}\over{r+\alpha}}\right)dt^2 - \left({{r + \alpha}\over{r-\alpha}}\right)dr^2 - \left(1+ {{\alpha}\over{r}}\right)r^2( d\theta^2 + \sin^2\theta\, d\phi^2).\eqno (6)$$
\medskip
\noindent Line element (6) clearly satisfies the Minkowski boundary requirements that 
as $r \to \infty$ then $g_{11} = 1,\ g_{22} -1,\ g_{33} = -r^2,\ g_{44}= -
r^2\sin^2\theta$ for the $t,\ r,\ \theta, \ \phi$ coordinates. Fock (1959$,$ pp. 194-203) uses (6) to obtain the usual prediction for the procession of a planetary orbit about the Sun$,$ and the ``bending'' of light rays as they pass near to the Sun. Moreover$,$ analysis of (6) also yields a D-type singularity and an event horizon surrounding this singularity as 
 it is identified by these Fock coordinates.\parm    
\centerline{\bf Explicit Contradictions}\parm
  The physically described event horizons two way 
communication properties are considered as real physical properties.\par
{\leftskip=0.5in \rightskip=0.5in \noindent  $\{\b 7\}$ On the 
other hand$,$ the presence or absence of horizons does not depend on the choice 
of coordinates. . . . all observers$,$ using whatever coordinates they like$,$ 
agree on the existence and location of the surfaces across which two-way 
communication is impossible. $\{\b 7\}$ (Ohanian$,$ 1976$,$ p. 312)\pars . . . . horizons represent physical 
coordinate-independent properties of spacetime$,$ . . . (Ohanian and Ruffini$,$ 1994$,$ p. 449). \par
}\par   
Relative to a freely falling explorer$,$ the following is claimed to be one of the major features for an event horizon.\par
{\leftskip=0.5in \rightskip=0.5in \noindent $\{\b 8\}$ Of course$,$ proper time is the relevant quantity for the explorer's heart-beat and health. No coordinate system has the power to prevent him from reaching $r = r_s.$ $\{\b 8\}$(Misner et.el. 1973$,$ p. 821)\par \noindent $\{\b 9\}$ In the rest frame of a falling astronaut the amount of proper time needed to enter a black hole and crash into the singularity is not only finite$,$ but also quite short. $\{\b 9\}$ (Ohanian and Ruffini$,$ 1994$,$ p. 448) \par}\par   
The EFT and the KST are not defined at $r=r_s,$ that is: they are piecewise defined. From a purely mathematical viewpoint$,$ the line elements obtained should also be considered as piecewise defined although it is claimed that applying the continuity argument to the line elements they are now defined at 
$r =r_s.$ For all that follows$,$ consider the following transformation that is not defined at $r =0.25$ where$,$ according to $\{\b 6\},$  $r$ ``must'' be regarded as the radial coordinate. Further$,$ $\rho \geq 0$ and a fixed mass is considered as concentrated at a point and mass ``units'' are utilized such that $r_s = 1.$
$$\rho =4r^2/\vert 4r-1\vert,\ r \not= 0.25.               \eqno (7)$$
The metric coefficients for the line element obtained from (7) are$,$ for $r \not= 0.25,$
$$g_{11} = {{4r^2- \vert 4r-1 \vert}\over{4r^2}}, \ g_{22} = {{-256r^4(-\vert 4r-1 \vert +2r({\rm signum}(4r-1)))^2}\over{(4r-1)^4(4r^2 -\vert 4r-1\vert)}},$$
 $$ g_{33} = {{-16r^4}\over{(4r-1)^2}} = {{-16 r^2}\over{(4r-1)^2}}r^2,\ g_{44} = g_{33}\sin^2\theta.\eqno (7a)$$\par
 According to the rules$,$ these metric coefficients satisfy the 
Einstein-Hilbert equation for the regions $0 <r <0.25$ and $0.25 < r.$ Note that for the region $r > 0.25$ there is no spacetime signature change. However$,$ there is such a signature change within the region $0 <r <(\sqrt 2 -1)/2.$ Our analysis will mainly be concerned with the behavior in the vacuum region $r > 0.25.$  When the absolute value and signum operators are eliminated from the metric coefficients$,$ it appears necessary to present the line element in two forms$,$ one for $r >0.25$ and one for $0<r<0.25.$ The $g_{11}$ has two distinct forms that are, of course, equal and continuous at $r=0.25.$ The expressions for $g_{33}$ and $g_{44}$ are identical. The expressions for the two $g_{22}$ are different in that one is multiplied by the extra factor $f(r)= (2r-1)^2/(4r^2+4r-1),$ but they are both not defined at $r=0.25.$ Indeed,
$$g_{22} = -256{{r^4}\over{(4r-1)^2}},\ r > 0.25$$
$$g_{22} = -256{{r^4}\over{(4r-1)^2}}f(r),\ 0 <r< 0.25,\ 
\lim_{r \to 0.25}f(r) = f(0.25) = 1. \eqno (7b)$$\par
It is obvious from (7b) that the asymptotic behavior about $r =0.25$ for these two different $g_{22}$ is essentially identical and, indeed, very well behaved. This is not the case with the Schwarzschild line element (5) which is not well behavior in neighborhoods about $\rho = r_s$ where it has divergent asymptotic behavior. However$,$ stated rules such as $\{\b 1\}$ and $\{\b 3\}$ are not relative to a specific form in which one might express a transformed line element. Indeed$,$ we are told that the behavior of the``geometry'' as described by the Riemann-tensor components$,$ and the describable behavior of particles or photons is the correct method to apply in analyzing line elements. \par    
In order to determine the local ``geometry'' of space time in the simplest possible manner$,$ $\{\b 5\}$ is applied. To properly calculate the ``static'' Riemann-tensor components in geodesic coordinates$,$ the software GRTensorII (Musgrave$,$ Pollney$,$ Lake$,$ 1997) as developed at Queens University$,$ Kingston OT Canada$,$ is utilized.  For $r >0.25$$,$ the following components were calculated using GRTensorII by a member of the Physics Dept. at Queens University and verified by this author.
 $$R_{\hat t \hat r \hat t \hat r} = {{(4r-1)^3}\over{64r^6}},\ R_{\hat \theta \hat \phi \hat \theta \hat \phi} = -{{(4r-1)^3}\over{64r^6}},$$ $$R_{\hat r \hat \theta \hat r \hat \theta}= R_{\hat r \hat \phi \hat r \hat \phi}={{(4r-1)^3}\over{128r^6}},\ R_{\hat t \hat \theta \hat t \hat \theta}=R_{\hat t \hat \phi \hat t \hat \phi}=-{{(4r-1)^3}\over{128r^6}}.$$
For the case where $0<r < 0.25,$ these same components are 
$$R_{\hat t \hat r \hat t \hat r} = -{{(4r-1)^3}\over{64r^6}},\ R_{\hat \theta \hat \phi \hat \theta \hat \phi} = {{(4r-1)^3}\over{64r^6}},$$ $$R_{\hat r \hat \theta \hat r \hat \theta}= R_{\hat r \hat \phi \hat r \hat \phi}=-{{(4r-1)^3}\over{128r^6}},\ R_{\hat t \hat \theta \hat t \hat \theta}=R_{\hat t \hat \phi \hat t \hat \phi}={{(4r-1)^3}\over{128r^6}}.$$
The first obvious and remarkable aspect of these static components is that they all converge to 0 as $r \to 0.25$ and can be so defined in order to maintain continuity at $r = 0.25.$ The next obvious aspect is that the components for $r <0.25$ all diverge as $r \to 0+.$ Although one could analyze these components for $r < 0.25,$ our interest is to investigate physical behavior for $r\geq 0.25.$ For example$,$ these static components have no other zero values for any $r > 0.25$ and they are maximal$,$ in the absolute sense$,$ at $r = 0.5.$ [Note: An additional factor called the Lorentz boost does not appear to alter the physical behavior associated with these static components.]\par
For all of the following analysis$,$ it is assumed that we are concerned with a configuration with material radius $r$ such that $0.25 \leq r < 0.5.$ To find the infinite redshift surfaces$,$ consider the argument of Ohanian and Ruffini (1994$,$ p. 439). We assume that $dr=d\theta=d\phi = 0$ and since we are measuring in light unites$,$ the proper time expression $d\tau^2= ds^2.$ This yields that 
$$d\tau = \pm{{2r-1}\over{2r}}dt.\eqno(8)$$
The exact same argument given by Ohanian and Ruffini in terms of the ``time'' dilation of $t$ yields that $r = 0.5$ is an infinite redshift surface. 
But$,$ of course$,$ this is in terms of the $(r,t)$ coordinates. Although in accordance with the rules for interpreting coordinate transformations$,$ the question should not be investigated$,$ we can ask whether $r$ should be considered as a true ``distance'' measure. 
For large values of $\rho$ and $r,$ say$,$ $1,000$ and beyond$,$ the $\rho$ and $r$ are approximately linearly related. Such a relation appears always to be acceptable. Further$,$ from $r = 10$ and beyond the difference 
$r- \rho < .2565$ and converges to $.25$.   
According to quotations $\{\b 1\}$$,$ $\{\b 2\}$ and $\{\b 3\}$ a physically described property such as infinite redshift of a light signal should not depend upon the coordinate measure $\rho$ or $r$. An observer at a large  $\rho$ and $r$ ``distance,'' that when compared are ``approximately'' the same and linearly related$,$ would notice the increasing redshift for either coordinate representation $(r,t)$ or $(\rho,t).$ If such a physical description does not depend upon an individual's defined $\rho$ coordinate$,$ then it cannot depend upon an individual's corresponding $r$ coordinate according to the rules. \par    
One of the most basic coordinate methods used to locate an event horizon is to analyze the ``one-way membrane'' concept relative to the geodesic path of light.   
Consider the geodesic of a ``light signal propagating in the radial direction'' (Ohanian and Ruffini$,$ 1994$,$ p.445). Hence$,$ let $ds=0$$,$ where $d\theta = d\phi = 0.$  In $(r,t)$ coordinates$,$ one obtains 
$${{dr}\over{dt}} = \pm{{(2r-1)(4r-1)^2}\over{32r^3}}.\eqno (9)$$
As stated in $\{\b 7\}$$,$ the existence and location of event horizons is not dependent upon a coordinate system. Further$,$ as argued above$,$ whether we use $r$ or $\rho$ measures should not make any difference in the gleaned physical behavior at $r=0.5$. The analysis used sets $dr/dt = 0$. In this case$,$ we have two possibilities$,$ $r = 0.5$ or $r = 0.25$. Unfortunately$,$ referring back to the metric coefficients$,$ one notices that there is no signature change
within the region  $r > 0.25.$ Analysis of the light-cones shows there is no alteration in their orientation relative to the surface $r = 0.5,$ and$,$ hence$,$ this surface is not an event horizon as described by the $(r,t)$ system. But$,$ it is an infinite redshift surface. This contradicts the requirement that for the Schwarzschild configuration being considered both of these surfaces should have the physical property of coinciding. [Indeed$,$ note that for $r=0.5,$ we have that $\rho = 1$; the surface for which it is claimed they coincide.] If one wishes to analysis the surface $r = 0.25,$ then it is also not an event horizon.\par     
Technically$,$ comparing this coordinate system with  another system in order to ``explain'' this apparent contradiction is forbidden for it would assume that the methods used are consistent. But$,$ an argument that might eliminate this contradiction$,$ using this forbidden method of comparison$,$ is that you should not use $r$ as a ``distance'' coordinate although the general rules stated in $\{\b 1\}$ - $\{\b 3\}$ don't include such an ad hoc requirement.   
Indeed$,$ the KST suffers from a ``physical'' difficulty when its coordinates are considered as ``real'' measures. If the above arguments relative to $(r,t)$ are not sufficient$,$ then one cannot argue for any physical concept using the KST $(u,v)$ coordinates.  Please note that$,$ in the KST $(u,v),$ $v$ behaves like time and $u$ like space measurements$,$ we are told. Indeed$,$ the entire discussion on page 453 of Ohanian and Ruffini (1994) treats $v$ as if it is ``time'' with all the intuitive time properties. We don't have an Einstein-Rosen bridge unless 
``[A]t the time $v=-3/2$,. . .'' there is  positive and {\it negative} value for $u$ that are obtained by intersection with a ``second $r = 0$'' branch$,$ a branch that does not exist in the Schwarzschild coordinates$,$ and ``At the time $\nu = -1$$,$ both of these singularities coalesce'' into a single point. Indeed$,$ it is stated  that as $u \to +\infty$ ``we have a space that is asymptotically flat. . . .'' and this should correspond to the same physical space as when $r \to +\infty.$  But if you let $r= 1.4$ and $t \to +\infty$$,$ then $u \to +\infty$ and the correspondence fails. \par    
Using statements $\{\b 4\}$ and $\{\b 8\},$ we show that statement $\{\b 9 \},$ relative to dustlike test particles$,$  is essentially contradicted by coordinate transformation (7).
The method in Rindler (1977$,$ p. 152) is applied. In order to solve the
free-fall geodesic equation for a material particle$,$ the first constant of integration is obtained by considering the particle at an extreme distance from
the center of attraction. In this case$,$ the space is approximately flat and$,$ in terms of proper time$,$ $dr/d\tau \approx 0 \approx dr/dt.$ Using the model that $\lim_{r\to\infty}dr/d\tau = \lim_{r\to\infty}dr/dt=0,$ the exact geodesic equations  of motion$,$ in proper time$,$ are
$${{dr}\over{d\tau}}=\pm {{(4r-1)^{5/2}}\over{16r^2(2r-1)}}.\eqno (10)$$ 
The free-fall geodesic equation for a material particle in terms of ``elapsed proper time'' $\tau$ and
for $r > 0.25,$ where the observation of its ``fall'' begins at
$r = a > 1/2,$ is the continuous function\par
$$\tau = \tau(r) +\tau(a) -4/3 + (4/3 -2\tau(r)){\rm H}(r- 1/2),$$
$$\tau(x)= {{16x^3}\over{3(4x-1)^{3/2}}},\eqno (11)$$
where H is the Heaviside (unit) step function.\par
Since this is a differential equation model$,$ this model would only apply$,$ in this exact manner$,$ to infinitesimal particles. Hence$,$ when one discusses the behavior of actual natural objects$,$ the equations only give approximate behavior and the description used should be somewhat general in character.
This discussion is only relative to the region $r>0.25.$ The term ``time'' refers to proper time and terms relative to the ``speed'' of motion refer  to proper speed. From a far away location$,$ the material particle starts slowly on its journey towards the center of attraction. The speed  increases until the particle reaches the infinite redshift surface. It passes this surface at its greatest speed and this corresponds to where the gravitational effects$,$ for this region$,$ are maximal. Once it passes this surface$,$ the speed decreases. 
Assume that specific Schwarzschild configuration with material ``radius'' $r = 0.25.$ [Note: Notwithstanding Rindler's comments relative to such a ``radial'' measure$,$ there is in the next section a discussion indicating what would occur if such a measure is rejected.] In order not to introduce a physical concept of ``infinity'' into this description$,$ one can proceed as follows: Starting from {\it any} specific radial position exterior to the infinite redshift surface (say $a = 1,000$) this real material particle takes a finite amount of time ($\approx 21,089$) to reach the infinite redshift surface. Consider any  natural number $n.$ The particle will require more than $na$ proper time units before it can ever reach the surface of the configuration.\par
  The above description contradicts 
quotation $\{\b 9\}$ and also contradicts a description for the behavior of the objects speed as represented by the differential equation on page 448 of Ohanian and Ruffini (1994). This physical description comes from an assurance by Rindler (1977$,$ p. 153) that by ``continuity it [the material  particle's geodesic] evidently represents one path.'' Further$,$ the  description in the above paragraph is in terms of physical concepts that are but gleaned from the coordinate system used.  
Note that the stated meaning of the geodesic path of a particle implies that if the particle like behavior of a material is used as a means to investigate the gravitational collapse of a Schwarzschild configuration within such a gravitational field$,$ then after the exterior surface of the configuration collapses through the infinite redshift surface$,$ the collapse could essentially stop and no actual ``black hole'' formed although a ``black hole'' would ``appear'' to exist under the usual observations. Such an appearance is due to the behavior of an infinite redshift surface$,$ which is not an event horizon$,$ and what are the maximal gravitational effects within this region. \parm
\centerline{\bf Countering These Contradictions}\parm
There are additional procedures that one might use in an attempt to counter these contradictions. Although it is not relevant to the contradictions presented here$,$ if one calculates the ``proper distance'' $d(r)$ (Misner$,$ et. al. 1973$,$ p. 824)$,$ taking into account the increasing or decreasing nature of ``proper velocity$,$'' then it follows that as $r \to 0.25+$ the distance $d(r) \to \infty.$ Hence$,$ in physical terms$,$ the proper distance becomes exceedingly great after the particle passes through the infinite redshift surface. This can be explained as simply the result of using proper time and proper speed to calculate the proper distance. Of course$,$ it could also be but another ``physical'' contradiction. Moreover$,$ one might note that as $r \to 0.25+$ then $\rho \to +\infty.$  However$,$ is the particle behavior thus far described in this section any less strange than that which occurs in the KST$,$ where as $r \to r_s$ then the time $t$ varies from $-\infty$ to $+\infty.$ Should an attempt be made to embed$,$ informally or rigorously$,$ this new coordinate system into the KST system? This would amount to considering other types of transformations that would ``paste'' portions of the manifold obtained from these new coordinates onto the KST manifold. Once again$,$ this is entirely unacceptable for it assumes the logical consistency of the GR methods used to argue for the KST physical scenarios and that the KST scenarios are the only correct ones. If one were to ignore this prohibition$,$ then the simplest identity transformation (i.e. the ``r'' in this new coordinate system corresponds to the ``$\rho$'' used in the KST) shows that this new coordinate system does completely cover the KST system and leads to the conclusion that the region of difficulty between the $r = 0.25$ surface and infinite red shift surface will be$,$ in the KST system$,$ a region where $u \to +\infty.$ This also would tend to lead to the rejection of the ``radius'' of the configuration as being measured by $r=0.25$ since such a measure corresponds to the Schwarzschild radius of $\rho = \infty.$ Using this ``forbidden'' transformation comparison approach or an approach that claims that the ``radius'' measurement of $r = 0.25$ has no physical meaning, one might claim that this is why the particle ``never'' reaches the $r = 0.25$ surface since the particle is actually ``moving towards'' totally flat space. However$,$ this conclusion would contradict the basic interpretation that a particle's continuous geodesic represents a free ``fall'' within a gravitational field.\par
The reason this previous approach is forbidden is a purely logical one. It is well-known
that if you do not add to or adjust the hypotheses used that the following is a correct logical argument. Consider statements P$,$ Q written in the appropriate manner and using only the predicates associated with GR. Suppose that we have logically established that the statement ``P and not P'' holds. Then we know that$,$ in general$,$ the statement ``P and not P implies Q'' holds logically. Hence$,$ the statement ``Q'' holds. But ``Q'' can be \underbar{any statement whatsoever.} \par   
The easiest method that could be used to counter these results is to simply give an ad hoc rejection based upon the ``obviously absurd'' physical predictions. Such coordinate system interpretations as here discussed would be called ``pathological''
by Misner$,$ Thorne and Wheeler (1973$,$ p. 934). However$,$ applying the Ferris comment$,$ the correct term 
is ``contradictory'' and this leads to all of the logically unacceptable difficulties inconsistency entails. In general$,$ these contradictions can be avoided in the same manner as is done within informal mathematics. As previously outlined$,$ informal mathematics is inconsistent if one is not careful as to how one defines entities. Mathematicians have learned how to do this by ``staying away'' as far as possible$,$ so to speak$,$ from such definitions. A similar Bergmann (1965) ad hoc philosophy of science could be applied. ``There exists a subset of physical variables$,$ the `observables,' whose values are independent of the choice of the  coordinate  system employed. Thus$,$ any relationship between observables is `meaningful,' and$,$ conversely$,$ these are the only relationships that are legitimate'' (p. 346). But$,$ using the Bergmann philosophy$,$ then the entire black hole scenario would be rejected. The arguments given in this paper indicate that it may be the methods used to glean physical behavior from coordinate system relationships$,$ using an inappropriate mathematical structure or not using a preferred coordinate system that might be the cause for these contradictions. Indeed$,$ the very Einstein-Hilbert equation with its correspondence between the energy-momentum tensor and Riemannian geometry could be totally in error. There are many present day theories that might be used to replace the Einstein theory. For example$,$ the very basic concepts of black hole theory need not be abandoned if these concepts are obtained by different means.
Such a means and a corresponding theory does exist (Herrmann$,$ 1994b). \parm
\centerline{\bf Conclusions}\parm
Relative to cosmology$,$ these research findings are of significance if the calculations are found to be valid. These findings would  affect the relationship between GR cosmological models for how our universe develops$,$ and the methods utilized to obtain such models. Clearly$,$ a significant scientific concept associated with cosmological models is strict consistency relative to scientific logic. Such cosmological models would be worthless if contradictions should occur. The results of this research should be kept in mind when such models are constructed using scientific logic applied to secular theories$,$ theories which may rely upon methods or hypotheses 
that lead to inconsistencies. \parm
\centerline{\bf References}\parm 

\id{D}ingle H$,$ 1950. The Special Theory of Relativity$,$ Methane's Monographs on Physical Subjects$,$ London.
\id{E}{\accent "7F o}tv{\accent "7F o}s$,$ R. V. 1889. {\accent "7F U}ber die Anziehung der Erde auf Verschiedene. Nath. Natruw. Ber. aus Ungram 8: 65-68.

\id{F}erris$,$ T. 1979. The Red Limit$,$ Bantam books$,$ New York.

\id{F}ock$,$ V. 1959. The theory of Space Time and Gravitation$,$ Pergamon Press New York.

\id{H}errmann$,$ R. A. 1998. Newton's Second law holds in normed linear spaces. Far East J. Appl. Math. 2(3):183-190. [Note: There is a major typographical error on page 187. The lower case "m" in the last line of (4) should be an upper case "M."] 
\id{H}errmann$,$ R. A. 1994a. Solution to the General Grand Unification Problem. 
$<$http://xxx.lanl.gov/abs/astro-ph/9903110$>$
\id{H}errmann$,$ R. A. 1994b. Einstein Corrected.\hfill\break
$<$http://www.serve.com/herrmann/books.htm$>$
\id{L}awden$,$ D. F. 1982. An introduction to tensor calculus$,$ relativity and cosmology. John Wiley \& Sons$,$ New York.
\id{L}ema{\accent 94 i}tre$,$ G. 1933. L'univers en expansion$,$ Ann. Soc. Sci. Bruxelles I A53$,$ 51-85.
\id{M}isner$,$ C. W.$,$ K. S. Thorne and J. A. Wheeler. 1973. Gravitation. Freeman$,$ San Fran.
\id{M}usgrave$,$ P.$,$ D. Pollney and K. Lake. 1997. GRTensorII Version 1.64 (R4). Queens University$,$ Canada$,$
$<$http://astro.queensu.ca/$\sim$grtensor/$>$ 
\id{O}hanian$,$ H. 1976. Gravitation and Spacetime. W. W. Norton Co.$,$ New York.
\id{O}hanian$,$ H. and R. Ruffini. 1994. Gravitation and Spacetime. W. W. Norton Co.$,$ New York.
\id{R}icci-Curbastro and Levi-Civita. 1901. M\'ethods de calcul diff\'erential absolute leurs applications. Math Ann 54$,$ p. 125-201$,$ 608.

\id{R}indler$,$ W. 1977. Essential Relativity$,$ Springer-Verlag$,$ New York.

\id{R}oll$,$ P. G.$,$ R. Krotkov$,$ and R. H. Dicke. 1964. The equivalence of inertial and passive gravitational mass. Ann. Phys. (U. S. A) 26: 442-517.

\id{S}chwarzschild K. 1916. {\accent "7F U}nber das Gravitational eines Massenpunktes nach der Einsteinschen Theorie$,$ Sitzber. Deut. Akad. Wiss. Berlin$,$ Kl. Math.-Phys. Tech.$,$ 189-196. 
\id{S}hapiro$,$ S. L. and S. A. Teukolsky$.$ Formation of Naked Singularities: The Violation of Cosmic Censorship$,$ Phy. Rev. Ltr. 66(8): 994-997.\par\medskip

\end